\documentclass[3p,times]{elsarticle}
\usepackage{amsfonts}
\usepackage{multirow}
\usepackage{mathrsfs}
\usepackage{graphicx}
\usepackage{amsmath}
\usepackage{amssymb}
\usepackage{bm}
\usepackage{bbm}
\usepackage{color}
\usepackage{slashed}
\usepackage{diagbox}
\newcommand{\beq}{\begin{equation}}
\newcommand{\eeq}{\end{equation}}
\newcommand{\bqa}{\begin{eqnarray}}
\newcommand{\eqa}{\end{eqnarray}}

\def\Slash#1{{#1\!\!\!\slash}}
\newcommand{\nn}{\nonumber}
\begin{document}
%%%%%%%%%%%%%%%%%%%%%%%%%%%%%%%%%%%%%%%%%%%%%%%%%%%%%%%%%%%%%%%%%%%%%%%%%%%%%%
\title{\mbox{}\\[10pt]
Next-to-leading-order QCD corrections to $e^+e^-\to H+\gamma$}

%%%%%%%%%%%%%%%%%%%%%%%%%%%%%%%%%%%%%%%%%%%%%%%%%%%%%%%%%%%%%%%%%%%%%%%%%%%%%%
\author{Wen-Long Sang~\footnote{wlsang@ihep.ac.cn}}
 \address{School of Physical Science and Technology, Southwest University, Chongqing 400700, China\vspace{0.2cm}}

\author{Wen Chen~\footnote{chenwen@ihep.ac.cn}}
\address{Institute of High Energy Physics and Theoretical Physics Center for
Science Facilities, Chinese Academy of
Sciences, Beijing 100049, China\vspace{0.2cm}}
\address{School of Physics, University of Chinese Academy of Sciences, Beijing 100049, China\vspace{0.2cm}}

\author{Feng Feng~\footnote{F.Feng@outlook.com}}
\address{China University of Mining and Technology, Beijing 100083, China\vspace{0.2cm}}
\address{Institute of High Energy Physics and Theoretical Physics Center for
Science Facilities, Chinese Academy of
Sciences, Beijing 100049, China\vspace{0.2cm}}

\author{Yu Jia~\footnote{jiay@ihep.ac.cn}}
\address{Institute of High Energy Physics and Theoretical Physics Center for
Science Facilities, Chinese Academy of
Sciences, Beijing 100049, China\vspace{0.2cm}}
\address{School of Physics, University of Chinese Academy of Sciences, Beijing 100049, China\vspace{0.2cm}}
\address{Center for High Energy Physics, Peking University, Beijing 100871,
China\vspace{0.2cm}}

\author{Qing-Feng Sun~\footnote{qfsun@mail.ustc.edu.cn}}
\address{Department of Modern Physics, University of Science and Technology of China, Hefei, Anhui 230026,
China\vspace{0.2cm}}
\address{Institute of High Energy Physics and Theoretical Physics Center for
 Science Facilities, Chinese Academy of
 Sciences, Beijing 100049, China\vspace{0.2cm}}

\date{\today}

%%%%%%%%%%%%%%%%%%%%%%%%%%%%%%%%%%%%%%%%%%%%%%%%%%%%%%%%%%%%%%%%%%%%%%%%%%%%%%
\begin{abstract}
The associated production of Higgs boson with a hard photon at lepton collider, {\it i.e.}, $e^+e^-\to H\gamma$,
is known to bear a rather small cross section in Standard Model, and can serve as a sensitive probe for
the potential new physics signals. Similar to the loop-induced Higgs decay channels $H\to \gamma\gamma, Z\gamma$,
the $e^+e^-\to H\gamma$ process also starts at one-loop order provided that the tiny electron mass is neglected.
In this work, we calculate the next-to-leading-order (NLO) QCD corrections to this associated $H+\gamma$
production process, which mainly stem from the gluonic dressing to the top quark loop.
The QCD corrections are found to be rather modest at lower center-of-mass energy range ($\sqrt{s}<300$ GeV),
thus of negligible impact on Higgs factory such as CEPC. Nevertheless, when the energy is boosted to the ILC energy range
($\sqrt{s}\approx 400$ GeV), QCD corrections may enhance the leading-order cross section by $20\%$.
In any event, the $e^+e^-\to H\gamma$ process has a maximal production rate $\sigma_{\rm max}\approx 0.08$ fb
around $\sqrt{s}= 250$ GeV, thus CEPC turns out to be the best place to look for this rare Higgs production process.
In the high energy limit, the effect of NLO QCD corrections become completely negligible, which can be simply attributed to the different asymptotic scaling behaviors of the LO and NLO cross sections,
where the former exhibits a milder decrement $\propto 1/s$ , but the latter undergoes a much faster decrease $\propto 1/s^2$.
\end{abstract}

%%%%%%%%%%%%%%%%%%%%%%%%%%%%%%%%%%%%%%%%%%%%%%%%%%%%%%%%%%%%%%%%%%
\maketitle
\section{Introduction}

After the historical discovery of the $125$ GeV boson by the \textsf{ATLAS} and \textsf{CMS} collaborations at LHC in 2012~\cite{Aad:2012tfa,Chatrchyan:2012xdj}, a great amount of efforts have been devoted to unravelling its
nature. Numerous evidences have been accumulating to indicate that this new particle is just
the long-sought Higgs boson, which plays the pivotal role in mediating spontaneous electroweak symmetry breaking.
To date, the measured couplings between the Higgs boson and heavy fermions/gauge bosons
are compatible with the Standard Model (SM)
predictions within $10\%-20\%$ accuracy~\cite{Khachatryan:2016vau}.

One of the central goals of contemporary high-energy physics is to precisely nail down the properties of
the Higgs boson, and to search for the footprint of new physics in the Higgs sector.
In contrast to the hadron colliders that are plagued with enormous background events,
lepton colliders appear to be much more appealing options for conducting precision measurements on Higgs properties.
Three promising next-generation $e^+e^-$ colliders, International Linear Collider (ILC) in Japan~\cite{Baer:2013cma,Asner:2013psa}, Future Circular Collider (FCC-ee) at CERN~\cite{Gomez-Ceballos:2013zzn} (formerly called TLEP), and Circular Electron-Positron Collider (CEPC) in China~\cite{CEPC-SPPCStudyGroup:2015csa,CEPC-SPPCStudyGroup:2015esa}, have been proposed in recent years.

All these three $e^+e^-$ colliders plan to operate at center-of-mass (CM) energy around 250 GeV, where the dominant Higgs production channel is via the so-called Higgsstrahlung process, $e^+e^-\to H Z$. For this reason, these $e^+e^-$ machines
are collectively referred to as Higgs factory.
This golden process has been intensively studied theoretically in the past decades, {\it e.g.,}
the leading order (LO) prediction was first given in Refs.~\cite{Ellis:1975ap,Ioffe:1976sd,Bjorken:1977wg},
while the next-to-leading order (NLO) electroweak corrections were analyzed in Refs.~\cite{Fleischer:1982af,Kniehl:1991hk,Denner:1992bc}.
Very recently, the mixed electroweak-QCD next-to-next-to-leading order (NNLO)
corrections have also been investigated by two groups~\cite{Sun:2016bel,Gong:2016jys},
which yield the state-of-the-art prediction for $\sigma(HZ)$ about $230$ fb around $\sqrt{s}=240$ GeV.

The motif of this work is to study another type of Higgs production process at future lepton colliders,
the associated production of Higgs boson with a hard photon, that is, $e^+e^-\to H\gamma$. Owing to the
exceedingly small electron Yukawa coupling and the absence of $H\gamma\gamma$,
$H\gamma Z$ couplings at tree level, this process first arises at one-loop order in SM.
The LO prediction to $\sigma(H\gamma)$ in SM is available long ago~\cite{Barroso:1985et,Abbasabadi:1995rc,Djouadi:1996ws},
which turns out to be several orders-of-magnitude smaller than that of $\sigma(HZ)$ around the Higgs factory energy.

Due to the highly suppressed production rate predicted in SM, the discovery prospect of the
$e^+e^-\to H\gamma$ process appears to be rather obscure in the future $e^+e^-$ colliders.
On the other hand, this may turn into a virtue, since this rare Higgs production process can serve as a
sensitive probe for new physics search. Had the couplings between the Higgs boson and the gauge bosons/top quark
been notably modified by some beyond-SM models, or had some hypothesized
heavy charged particles been strongly coupled to the Higgs boson,
the production cross section for $\sigma(H\gamma)$ might be substantially enhanced relative to its SM value,
so that the $e^+e^-\to H\gamma$ process could even possibly be observed at Higgs factory.
The impact of possible new physics scenarios on this process has been extensively investigated
in literatures~\cite{Gounaris:1995mx,Mahanta:1997hg,Arhrib:2014pva,Hu:2014eia,
Ren:2015uka,Cao:2015iua,Li:2015kxc}.

Needless to say, an accurate SM account for the $e^+e^-\to H\gamma$ process is crucial and mandatory for confidently
interpreting the potential experimental signal in future.
The goal of this work is to conduct a detailed study on the NLO QCD corrections to this process in various energy range.
In a sense, the required two-loop calculation is similar to the previous works about NLO QCD corrections to $H\to \gamma\gamma$~\cite{Zheng:1990qa,Djouadi:1990aj,Dawson:1992cy,Djouadi:1993ji,Melnikov:1993tj,Inoue:1994jq,
Spira:1995rr,Fleischer:2004vb, Harlander:2005rq,Aglietti:2006tp} and
$H\to Z\gamma$~\cite{Spira:1991tj,Bonciani:2015eua,Gehrmann:2015dua}.
Nevertheless, the situation in our case is more involved than these Higgs decay processes,
due to the occurrence of a new energy scale, $\sqrt{s}$.
We are also interested in inferring how the effects of NLO QCD corrections vary with $\sqrt{s}$.

The rest of this paper is organized as follows. In Section~\ref{sec-theory},
we establish the notations and briefly describe the procedure of our NLO calculation.
Section~\ref{sec-results} is dedicated to presenting our numerical results.
Finally we present a summary and outlook in Section~\ref{sec-summary}.
For the convenience of the readers, the compact expressions for the LO amplitude
are collected in Appendix.

\section{Descriptions of the NLO calculation\label{sec-theory}}

Owing to the Lorentz covariance and electromagnetic current conservation,
the amplitude for $e^+e^-\to H\gamma$ can be decomposed into the linear combination of
four distinct Lorentz structures~\cite{Djouadi:1996ws}:
%------------------------
\bqa{\label{eq-amp-dec}}
%------------------------
{\cal M}=\sum_{i=1,2;\,\alpha=\pm}{C^\alpha_i \Lambda^\alpha_i },
%------------------------
\eqa
%------------------------
where $C_{1,2}^\pm$ are scalar coefficients, and
%------------------------
\begin{subequations}
%------------------------
{\label{eq-lor-4}}
%------------------------
\bqa
%------------------------
\Lambda_1^{\pm}=\bar{v}(p_+)(1\pm\gamma_5)(\Slash{\varepsilon}^{*}_\gamma p_\gamma\cdot p_--\Slash{p}_\gamma\varepsilon^{*}_\gamma\cdot p_-)u(p_-),\\
\Lambda_2^{\pm}=\bar{v}(p_+)(1\pm\gamma_5)(\Slash{\varepsilon}^{*}_\gamma p_\gamma\cdot p_+-\Slash{p}_\gamma\varepsilon^{*}_\gamma\cdot p_+)u(p_-),
%------------------------
\eqa
%------------------------
\end{subequations}
%------------------------
where $p_{+}$ and $p_{-}$ signify the momenta of the incoming positron and electron,
$p_\gamma$ and $p_H$ signify the momenta of the outgoing photon and Higgs boson,
and $\varepsilon_\gamma$ denotes the photon polarization vector.
For simplicity, we shall neglect the tiny electron mass throughout this work.

Substituting (\ref{eq-lor-4}) into (\ref{eq-amp-dec}), squaring, averaging upon the
$e^\pm$ spins
and summing over photon helicity, we can express the unpolarized  differential cross section
for $e^+e^-\to H\gamma$ as
%------------------------
\bqa{\label{eq-cs-f}}
%------------------------
\frac{d\sigma}{d\cos\theta}=\frac{1}{2s}\frac{s-m_H^2}{16\pi s}\frac{1}{4}\sum_{\rm pol}{\left|{\cal M}\right|^2}=\frac{s-m_H^2}{64\pi s}\bigg[
t^2\left(|C_1^+|^2+|C_1^-|^2\right)+u^2\left(|C_2^+|^2+|C_2^-|^2\right)
\bigg],
%------------------------
\eqa
%------------------------
where $\theta$ denotes the polar angle between the the outgoing Higgs and the incoming positron, and $m_H$ represents the Higgs boson mass. $s=(p_++p_-)^2$, $t=(p_H-p_+)^2$, and $u=(p_H-p_-)^2$ are standard Mandelstam's variables.

%-------------------------------------------------
\begin{figure}[ht]
\begin{center}
\includegraphics*[scale=1.0]{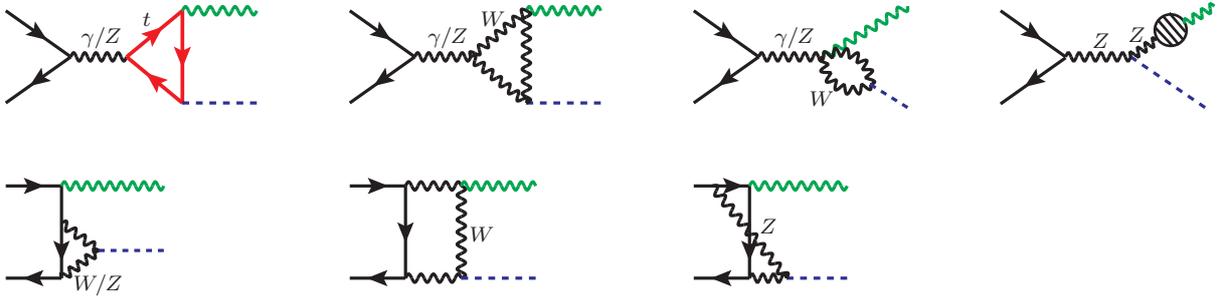}
\caption{Representative Feynman diagrams for $e^+e^-\to H\gamma$ at lowest order
in SM.
The shaded circle in the fourth diagram, which designates one-loop diagrams routed with
charge particles, mediates the mixing between $Z^0$ and $\gamma$.
}
\label{fig-feyn-LO}
\end{center}
\end{figure}
%-------------------------------------------------

This process first occurs at one-loop order in SM, with some typical diagrams shown in Figure~\ref{fig-feyn-LO}.
The calculation of the LO amplitude has been comprehensively described in Refs.~\cite{Barroso:1985et,Abbasabadi:1995rc,Djouadi:1996ws}.

To assess the impact of the NLO QCD corrections, it is convenient to expand the scalar form factors
$C_{1,2}^\pm$ in powers of the strong coupling constant:
%------------------------
\bqa{\label{eq-c-split}}
%------------------------
C_{1,2}^\pm=C_{1,2}^{\pm (0)}+\frac{\alpha_s}{\pi}C_{1,2}^{\pm (1)}+\cdots,
%------------------------
\eqa
%------------------------
where $C_{1,2}^{\pm (0)}$ designate the LO contributions, and $C_{1,2}^{\pm (1)}$ encode
the relative order-$\alpha_s$ corrections.

Substituting (\ref{eq-c-split}) back into (\ref{eq-cs-f}), employing $|C_{1,2}^{\pm}|^2\approx |C_{1,2}^{\pm(0)}|^2+
2 {\alpha_s\over \pi}\,{\rm Re}
\left[C_{1,2}^{\pm(0)} C_{1,2}^{\pm(1)*}\right]$, one then readily identifies the NLO QCD corrections
to the differential cross section. The LO expressions for $C_{1,2}^{\pm (0)}$ are well recorded in literatures~\cite{Barroso:1985et,Abbasabadi:1995rc,Djouadi:1996ws}.
The central challenge of this work is then to compute the four form factors $C_{1,2}^\pm$
through order $\alpha_s$.

In our calculation, we employ the Feynman gauge in both electroweak and QCD sectors. Moreover, we adopt
the dimensional regularization to regularize both UV and IR divergences. The LO and NLO Feynman diagrams and corresponding amplitudes are generated by \textsf{FeynArts}~\cite{Hahn:2000kx}.
We use the \textsf{Mathematica} packages \textsf{FeynCalc}~\cite{Mertig:1990an,Shtabovenko:2016sxi}/\textsf{FeynCalcFormLink}~\cite{Feng:2012tk}
to carry out the trace over Dirac/color matrices, and utilize the packages \textsf{Apart}~\cite{Feng:2012iq} (and another private code) and the \textsf{C++} package \textsf{FIRE}~\cite{Smirnov:2014hma} to carry out the
partial fraction together with the integration-by-parts (IBP) reduction for tensor integrals,
finally end up with a number of master integrals (MIs).

Similar to the loop-induced processes $H\to \gamma\gamma, Z\gamma$,
the LO amplitude for $e^+e^-\to H\gamma$ is also UV finite.
The corresponding $C_{1,2}^{\pm(0)}$ can then be expressed in terms of the
linear combinations of a number of one-loop Passarino-Veltman functions.
These form factors have been presented in this format in previous work~\cite{Djouadi:1996ws},
while retaining a non-vanishing electron mass to regulate the potential collinear divergence (which can arise from the last diagram in Figure~\ref{fig-feyn-LO}).
Of course, the ultimate amplitude is free from collinear singularity and one is allowed to take the smooth limit
$m_e\to 0$ in the end.
As stated before, we have set $m_e=0$ at the outset and used dimensional regularization
to regularize both UV and collinear divergences.
We feel that our procedure looks simpler both conceptually and technically.
For the sake of completeness, and for readers' convenience,
we present our condensed expressions for $C_{1,2}^{\pm(0)}$ in the Appendix.
The involved Passarino-Veltman functions can be worked out analytically,
or can be accurately evaluated by the packages
\textsf{LoopTools}~\cite{Hahn:1998yk} and \textsf{Collier}~\cite{Denner:2016kdg}
in a numerical manner.

%-------------------------------------------------
\begin{figure}[ht]
\begin{center}
\includegraphics*[scale=1.0]{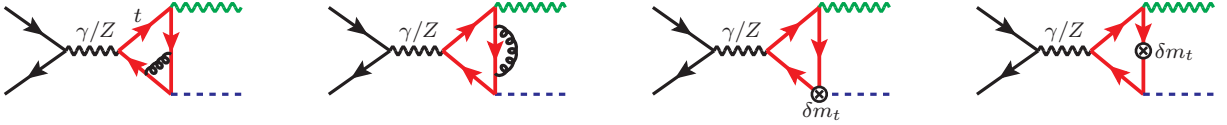}
\caption{Typical Feynman diagrams for the NLO QCD corrections to $e^+e^-\to H\gamma$.
The cap signifies the insertion of the top quark mass counterterm $\delta m_t$,
as given in (\ref{mt:mass:counterterm}).}
\label{fig-feyn-NLO}
\end{center}
\end{figure}
%-------------------------------------------------

We wish to describe more details about the calculation of the NLO QCD corrections.
At NLO, all the two-loop diagrams have simple $s$-channel topology, some of which
are illustrated in Figure~\ref{fig-feyn-NLO}.
They can simply be obtained by dressing the gluon to the top quark loop in all possible ways.
For simplicity, we have suppressed the contributions from five lighter quarks,
due to their much smaller Yukawa couplings.
Thanks to the simple $s$-channel topology, the amplitude assumes a factorized form:
%-------------------------
\bqa
%-------------------------
\label{eq-nlo-amp-factor}
%-------------------------
&& \mathcal{M}^{(1)} = e\bar{v}(p_+)\gamma_{\mu}\bigg(g_e^- P_- + g_e^+ P_+ \bigg)u(p_-)
{1\over s-M_Z^2}\mathcal{T}_{Z\gamma H}^{\mu\nu}\varepsilon_{\gamma,\nu}^{*}
+e\bar{v}(p_+)\gamma_{\mu} u(p_-){1\over s} \mathcal{T}_{\gamma \gamma H}^{\mu\nu}\varepsilon_{\gamma,\nu}^{*},
%-------------------------
\eqa
%-------------------------
where $P_\pm = \frac{1\pm \gamma_5}{2}$ are the chirality projection operators.
$g_f^-\equiv I^3_f/(s_Wc_W)-Q_f s_W/c_W$, and $g_f^+\equiv-Q_f s_W/c_W$, are the $Zf\bar{f}$ couplings for the left-handed and right-handed fermion $f$, respectively, where $Q_f$ and $I^3_f$ designate the corresponding electric charge and the third component of
the weak isospin of the fermion $f$. $c_W(s_W)$ stands for the cosine (sine) of the Weinberg angle.
The two terms in (\ref{eq-nlo-amp-factor}) stem from two distinct channels, $e^+e^-\to Z^* \to H\gamma$ and $e^+e^-\to \gamma^* \to H\gamma$, respectively.
$\mathcal{T}_{V \gamma H}^{\mu\nu}$ represent the corresponding tensor form factors
related to the $V\gamma H$ ($V=\gamma,Z$) vertex.

As mentioned before, at NLO, we only need consider the contribution to the ${V \gamma H}$ vertex from the top quark loop.
Lorentz covariance enables us to parameterize the vertex form factors
$\mathcal{T}_{V \gamma H}^{\mu\nu}$ ($V=\gamma,Z$) as
%-------------------------
\bqa\label{eq-T-ff}
%-------------------------
\mathcal{T}_{V \gamma H}^{\mu\nu}=T_{V,1}p_\gamma^\mu p_\gamma^\nu+T_{V,2}k^\mu k^\nu +
T_{V,3} p_\gamma^\mu k^\nu + T_{V,4} k^\mu p_\gamma^\nu + T_{V,5} p_\gamma\cdot k g^{\mu\nu}+ T_{V,6}\epsilon^{\mu\nu\alpha\beta} k_\alpha p_{\gamma,\beta},
%-------------------------
\eqa
%-------------------------
where $k=p_\gamma+p_H$ is the 4-momentum of the virtual gauge boson $V$, and $T_{V,i}$ ($i=1,\ldots,6$) are
various scalar form factors as the functions of $s$ and $m_H$.
Furry's theorem, equivalently the charge conjugation symmetry, forbids
the axial vector coupling of $Zt\bar{t}$ to yield a nonvanishing contribution (at least at this order),
consequently $T_{V,6}=0$.
Electromagnetic current conservation implies that $T_{V,2}=0$, and
$T_{V,3}=-T_{V,5}$. Furthermore, after substituting (\ref{eq-T-ff}) into (\ref{eq-nlo-amp-factor}), one
finds that $T_{V,1}$ and $T_{V,4}$ do not contribute to the amplitude.
Consequently, we just need to calculate a single scalar form factor $T_{V,5}$.
We emphasize that, the mere QCD renormalization required in this work is to
implement the top quark mass renormalization, with the mass counterterm:
%-------------------------
\beq\label{mt:mass:counterterm}
%-------------------------
\delta m_t = - \frac{3C_F}{4}\frac{\alpha_s}{\pi}\bigg[\frac{1}{\epsilon}+\frac{4}{3}+
\ln\frac{4\pi\mu^2}{m_t^2}-\gamma_E\bigg] m_t,
%-------------------------
\eeq
%-------------------------
where the spacetime dimensions $d=4-2\epsilon$, and
$m_t$ signifies the top quark pole mass.
The counterterm diagrams are represented by the last two diagrams
in Figure~\ref{fig-feyn-NLO}.

Having eliminated the UV divergences in $T_{V,5}$ ($V=Z,\gamma$) that enter the NLO amplitude
in (\ref{eq-nlo-amp-factor}),
after some straightforward manipulations,
one can identify the NLO coefficients $C_{1,2}^{\pm(1)}$ with
%-------------------------
\bqa\label{eq-c-1:decomposition}
%-------------------------
C_{1}^{\pm(1)}=C_{2}^{\pm(1)}=\frac{e}{2}g_e^\mp\frac{1}{s-M_Z^2}T_{Z,5} +\frac{e}{2s}T_{\gamma,5}
={e\over 2 s}\left(1+ g_e^\mp \kappa {s\over s-M_Z^2} \right) T_{\gamma,5},
%-------------------------
\eqa
%-------------------------
where $\kappa \equiv T_{Z,5}/T_{\gamma,5}= -\tfrac{g_t^+ + g_t^-}{2 Q_t}$.
It is easy to understand why $T_{Z,5}$ and $T_{\gamma,5}$ are proportional to each other,
since only the vector part of the $Z t\bar{t}$ coupling survives in the amplitude.
%-------------------
Plugging (\ref{eq-c-1:decomposition}) into (\ref{eq-c-split}) and (\ref{eq-cs-f}),
we then deduce the desired NLO QCD corrections to the differential cross section.

We pause here to briefly describe our numerical strategy of computing the NLO corrections. After IBP reduction,
we end up with about 34 two-loop MIs for $T_{\gamma,5}$. For each MI, we combine \textsf{FIESTA}~\cite{Smirnov:2015mct}/\textsf{CubPack}~\cite{CubPack} to perform the sector decomposition and the subsequent numerical integrations with quadruple precision.
We also utilize the Fortran package \textsf{Cuba}~\cite{Hahn:2004fe} to crosscheck the numerical integrations for most MIs.

It is worth mentioning that our two-loop calculation is quite similar to the corresponding calculation
for the rare decay process $H\to Z\gamma$. The NLO QCD corrections to that process were recently accomplished by two groups in an analytic fashion~\cite{Bonciani:2015eua,Gehrmann:2015dua}.
As a crosscheck, we have also revisited that process and found good agreement with the numerical results tabulated in~\cite{Bonciani:2015eua}. Having passed such a nontrivial test, we feel confident about the correctness of our
NLO QCD calculation for $e^+e^- \to H\gamma$.

In principle, our $T_{Z,5}$ can be obtained from its counterpart in $H\to Z\gamma$ by performing some sort of
analytic continuation. Unfortunately, the analytic expressions recorded in Refs.~\cite{Bonciani:2015eua,Gehrmann:2015dua}
involve rather lengthy and complicated special functions, rendering analytic continuation a rather nontrivial task.
Therefore, we are contented with only presenting the highly accurate numerical predictions in this Letter.

\section{Numerical results\label{sec-results}}

In order to make concrete predictions, we specify the following input parameters,
in accordance with the latest compilation of
the Particle Data Group~\cite{Olive:2016xmw}:
%-------------------------
\begin{eqnarray}
&&M_W=80.385\,{\rm GeV},\,\, M_Z=91.1876\,{\rm GeV},
\,\, m_t=174.2\,{\rm GeV},\,\,\nn\\
&& m_H=125.09\,{\rm GeV},\,\,\alpha=1/{137.035999}.
\end{eqnarray}
%-------------------------
In addition, the default renormalization scale $\mu=\sqrt{s}$ is assumed for the QCD running coupling constant $\alpha_s(\mu)$.
We take the initial value $\alpha_s^{(5)}(M_Z)=0.118$, and use the two-loop running formula to evolve to another
different energy scale. Practically, it is very convenient to utilize the package
\textsf{RunDec}~\cite{Chetyrkin:2000yt}, which has automated the evolution of the QCD running coupling and
taken care of the top quark threshold effect.

%-------------------------------------------------
\begin{figure}[ht]
\begin{center}
\includegraphics*[scale=0.6]{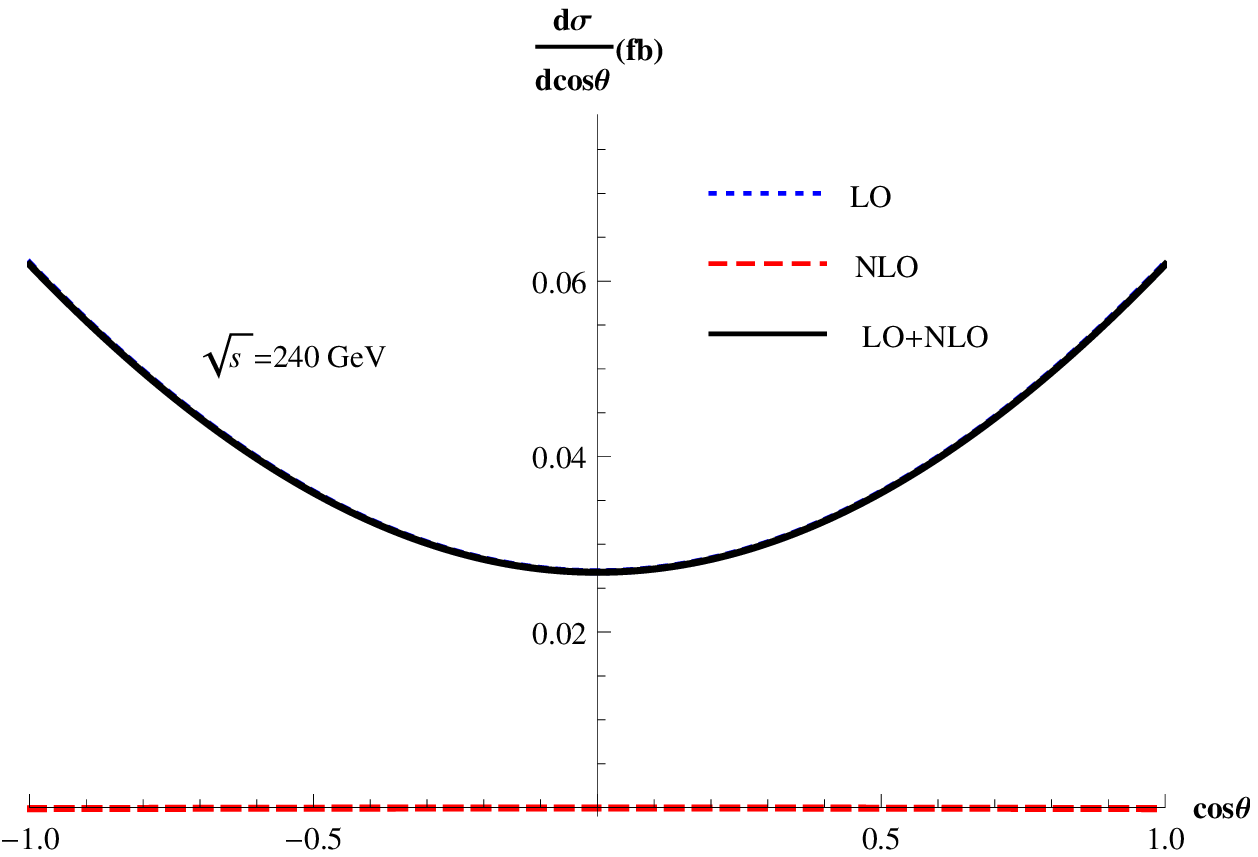}
\includegraphics*[scale=0.6]{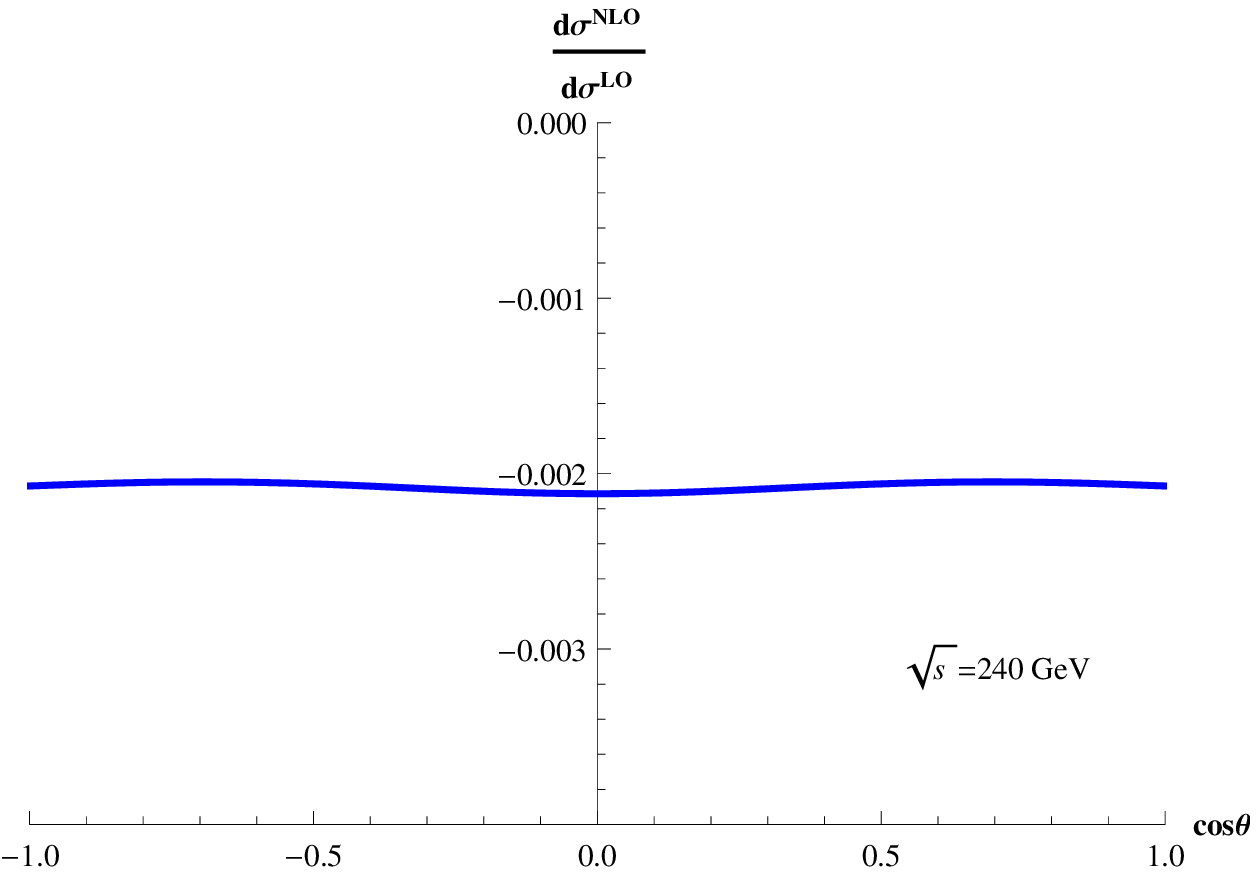}
\caption{Angular distributions of the Higgs boson in the $e^+e^-\to H\gamma$ process at $\sqrt{s}=240$ GeV.
The right panel embodies the relative magnitude of the NLO QCD corrections.}
\label{fig-angular-distribution-240}
\end{center}
\end{figure}
%-------------------------------------------------

%-------------------------------------------------
\begin{figure}[ht]
\begin{center}
\includegraphics*[scale=0.6]{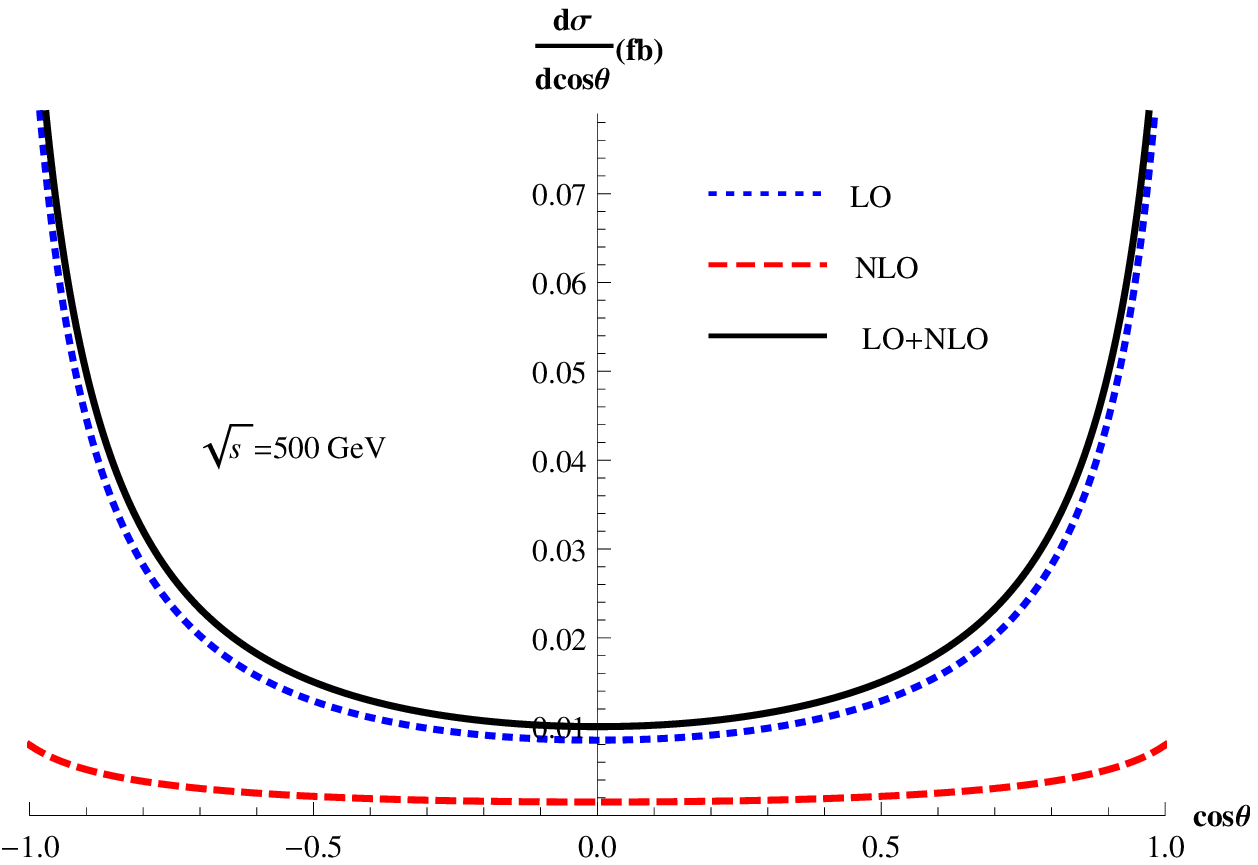}
\includegraphics*[scale=0.6]{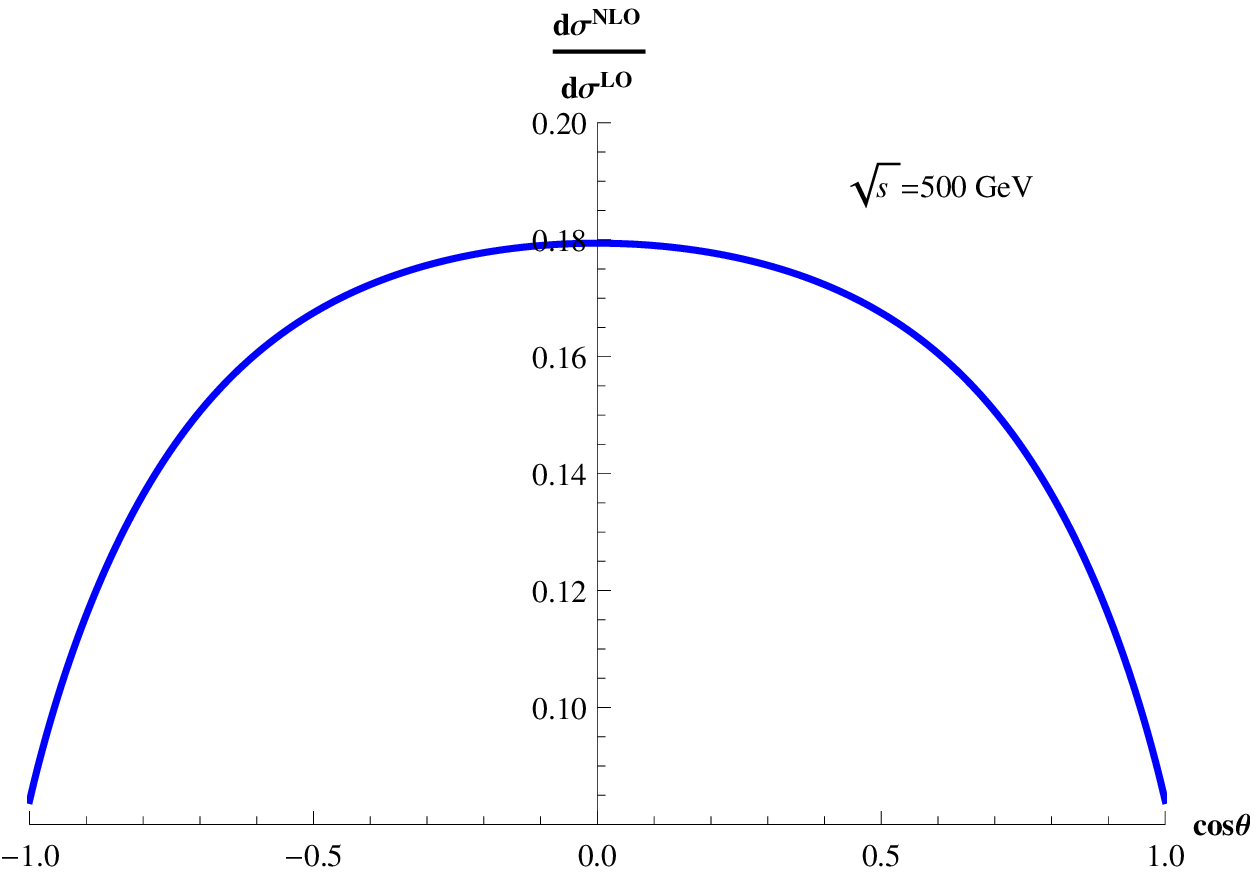}
\caption{Angular distributions of the Higgs boson in the $e^+e^-\to H\gamma$ process at $\sqrt{s}=500$ GeV.
The right panel signifies the relative magnitude of the NLO QCD corrections.}
\label{fig-angular-distribution-500}
\end{center}
\end{figure}
%-------------------------------------------------

Before proceeding into phenomenological discussions,
we should remind the readers that our LO predictions are subject to an intrinsic ambiguity
due to the choice of the electroweak coupling constants.
In this work, we have chosen to use the Thomson-limit value of QED coupling $\alpha(0)$.
Certainly, it is equally legitimate to adopt different values for $\alpha$, such as the widely-used
$\alpha_{G_\mu}$ and $\alpha(M_Z)$~\cite{Denner:1991kt}.
This uncertainty might be even more pronounced than the size of the NLO QCD corrections.
This ambiguity can only be lessened once one incorporates the full NLO weak corrections,
unfortunately which is currently unavailable.

In Figure~\ref{fig-angular-distribution-240} and \ref{fig-angular-distribution-500},
we plot the angular distributions of the Higgs boson at two benchmark CM energy points,
$\sqrt{s}=240$ GeV at CEPC, and 500 GeV at ILC.
As the CM energy increases, the angular distributions turn to be more and more
concentrated in the forward/backward directions.
As one can readily tell, the effect of NLO QCD corrections at CEPC energy,
which only constitutes $\sim 0.2\%$ of the LO contribution, is hardly discernible.
In the meanwhile, the impact of the NLO QCD corrections becomes clearly visible
at ILC energy range.

%-------------------------------------------------
\begin{figure}[hbt]
\begin{center}
\includegraphics*[scale=0.8]{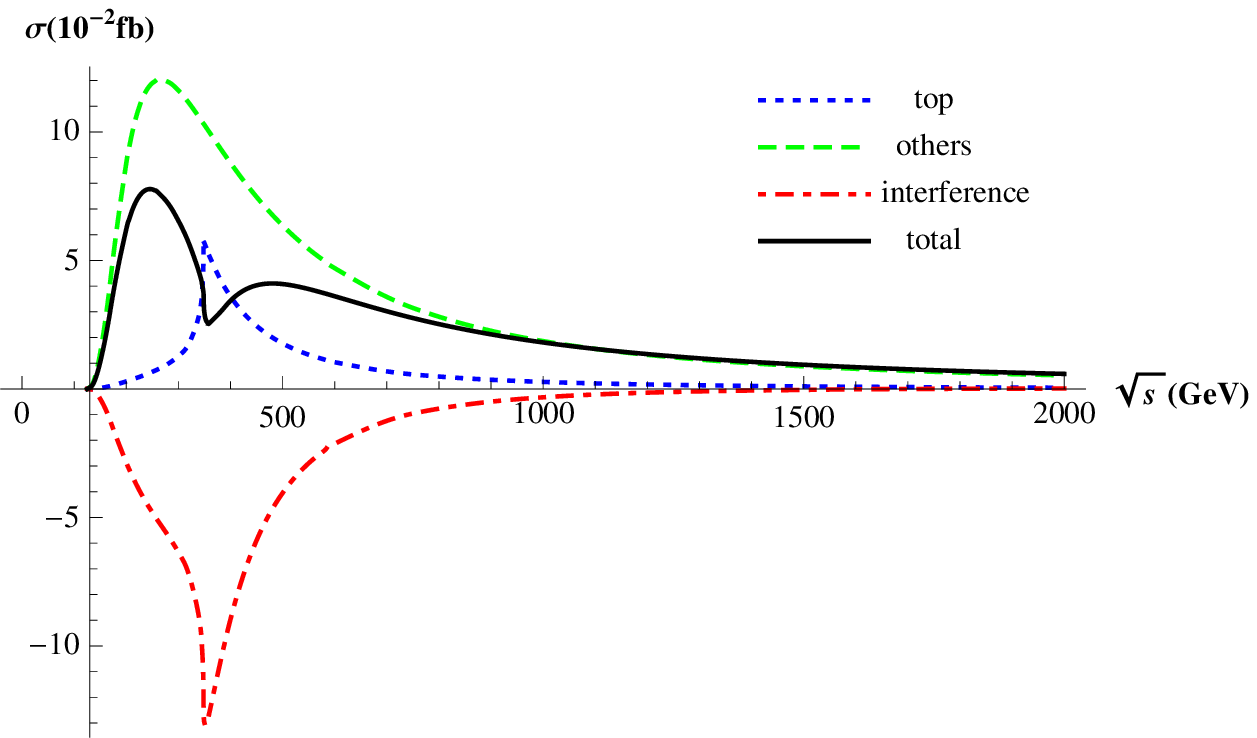}
\caption{The LO cross section as a function of $\sqrt{s}$ (the solid line). To trace the origin of the
nontrivial line shape, we deliberately isolate the contributions from two classes of diagrams.
The dotted, dashed and dot-dashed lines represent the contribution from diagrams involving
the top quark loop, that from all other diagrams involving weak gauge bosons in the loop,
and their interference,  respectively.}
\label{fig-cs-lo}
\end{center}
\end{figure}
%-------------------------------------------------

To ascertain the integrated cross section at each CM energy point, we utilize \textsf{Cuba}~\cite{Hahn:2004fe} to
carry out the angular integration numerically. In Figure~\ref{fig-cs-lo}, the variation of the LO cross section
with the CM energy is depicted over a wide range. There the line shape exhibits some interesting features.
Starting from the threshold $\sqrt{s}=m_H$, the cross section rises from zero due to the increment of the phase space.
The cross section reaches its first peak around $\sqrt{s}=245$ GeV, followed by a dip around $\sqrt{s}=359$ GeV,
then slowly bounces up and reaches the second lower peak centered at $\sqrt{s}=480$ GeV.
The cross section then decreases monotonically with increasing $\sqrt{s}$.

To understand the origin of the peak-dip-peak structure, we separate the contributions into two categories,
those from the diagrams involving the top quark loop and from the diagrams involving electroweak gauge bosons.
There arises a cusp from the former channel, exactly located at the $t\bar{t}$ threshold $\sqrt{s}=2m_t=348.4$ GeV.
From Figure~\ref{fig-cs-lo}, one observes that there arises a substantial destructive interference between these two channels,
whose effects reach the maximum at  $\sqrt{s}\approx 352$ GeV.
This destructive interference seems to be largely responsible for the global minimum of the net LO line shape.
It is interesting to note that the location of this dip deviates from the $t\bar{t}$ threshold by 10 GeV.

We wish to understand the asymptotic behavior of $\sigma(H\gamma)$ in the high energy limit. For this purpose,
it is instructive to temporarily digress into the asymptotic scaling of the total cross section
of a similar Higgs production channel, the Higgsstrahlung process.
It is well-known that $\sigma(HZ)$ scales as $1/s$ in the high energy limit~\cite{Fleischer:1982af,Kniehl:1991hk,Denner:1992bc}.
This scaling is saturated by the channel with a longitudinally-polarized $Z$,
while the polarized cross section for a transversely-polarized $Z$ declines at a much steeper rate,
$\propto 1/s^2$~\cite{Denner:1992bc,Sun:2016bel}.

Since the photon must be transversely polarized, one might naively anticipate that $\sigma(H\gamma)$
should decrease as $1/s^2$ asymptotically.
However, a careful numerical investigation reveals that this expectation is not true!
At sufficiently high energy, the cross section turns out to decline with a much slower pace,
 $\propto 1/s$. In addition to this power-law scaling, we have not observed any hint of
logarithmic enhancement, to a decent numerical accuracy.
Therefore, the $\ln^n s$ ($n=1,2$) terms, if exist, must be accompanied with the power-suppressed terms
($\propto 1/s^2$).

It is interesting to trace the origin for the $1/s$ scaling of $\sigma(H\gamma)$ at high energy.
Note there is one important difference between Higgsstrahlung and the process considered in this Letter,
{\it e.g.}, the former starts at tree level, but the latter starts at one loop order.
The $s$-channel diagrams in Figure~\ref{fig-feyn-LO} indeed only yield a suppressed contribution $\propto 1/s^2$,
modulo possible logarithms of $s$.
Nevertheless, the box diagrams in Figure~\ref{fig-feyn-LO},
which are responsible for the sharp concentration of the differential cross sections in the forward/backward zones,
will bring forth an enhancement factor $\propto {s\over M^2}$ (here $M$ refers to a generic mass scale
of order $M_{Z,W}$ or $M_H$) upon angular integration. A closer analytic examination also
supports the absence of logarithms of $s$ accompanying this leading-power scaling.

%-------------------------
\setlength{\tabcolsep}{2pt}
%-------------------------
\begin{table*}
\begin{center}
\begin{tabular}{|c|c|c|c|c|c|c|c|c|c|c|c|c|c|c|}
\hline
$\sqrt{s}({\rm GeV})$&150&200&220&240&250&270&290&310&330&340\\
\hline
{\rm $\sigma^{{\rm LO}}$ ($10^{-2}$ fb)}
&1.054&6.214&7.339&7.758
&$7.764$&$7.479$&$6.909$&$6.134$&$5.151$&4.522\\
\hline
$\widetilde{T}_{\gamma,5}$($10^{-2}$GeV$^{-1}$)
&-0.793&-0.378&-0.112&0.251
&0.485&1.12&2.11&3.90&8.16&14.26\\
%\hline
%$\tilde{T}_{Z,5}$($10^{-2}$GeV$^{-1}$)
%&0.290&0.138&0.041&-0.091
%&-0.177&-0.408&-0.772&-1.43&-2.98\\
\hline
{$\sigma^{{\rm NLO}}/\sigma^{{\rm LO}}$}
&0.56\%&0.30\%&0.09\%&-0.21\%
&-0.41\%&-0.96\%&-1.86\%&-3.45\%&-6.85\%&-10.59\%\\
\hline
%----------------------
\hline
$\sqrt{s}({\rm GeV})$&360&380&400&420&500&600&700&800&900&1000\\
\hline
{\rm $\sigma^{{\rm LO}}$ ($10^{-2}$ fb)}&2.570&2.977&3.433&3.763&4.079&3.604&3.018&2.518&2.118&1.801 \\
\hline
%$\tilde{T}_{\gamma,5}$($10^{-2}$GeV$^{-1}$)&$-11.6+16.6\,i$&$-13.4+9.81\,i$\;&$-9.65-1.29\,i$\;&$-6.31-3.21\,i$\;
% &$-4.45-3.48\,i$\;&$-3.35-3.38\,i$\;&$-2.13-3.00\,i$\;\\
$\widetilde{T}_{\gamma,5}$ ($10^{-2}$GeV$^{-1}$)\;& $-2.26$ &$-11.6$&$-13.4$\;&$-13.24$\;&$-9.65$\;&$-6.31$\;
 &$-4.45$\;&$-3.35$\;&$-2.63$\;&$-2.13$\;\\
&$+28.2\,i$&$+16.6\,i$&$+9.81\,i$\;&$+5.76\,i$\;&$-1.29\,i$\;&$-3.21\,i$\;
 &$-3.48\,i$\;&$-3.38\,i$\;&$-3.19$\,i&$-3.00\,i$\;\\
\hline
%$\tilde{T}_{Z,5}$($10^{-2}$GeV$^{-1}$)&$4.25-6.07\,i$&$4.89-3.65\,i$&$3.53+0.471\,i$&$2.31+1.17\,i$&$1.63+1.27\,i$&
%$1.22+1.23\,i$&$0.779+1.10\,i$\\
%$\tilde{T}_{Z,5} $($10^{-2}$GeV$^{-1}$)&$4.25$&$4.89$&$4.84$&$3.53$&$2.31$&$1.63$&
%$1.22$&$0.961$&$0.779$\\
%&$-6.07\,i$&$-3.65\,i$&$-2.12\,i$\;&$+0.471\,i$&$+1.17\,i$&$+1.27\,i$&
%$+1.23\,i$&$+1.17\,i$&$+1.10\,i$\\
%\hline
$\sigma^{{\rm NLO}}/\sigma^{{\rm LO}}$&-4.99\%&16.81\%&19.87\%&19.24\%&13.67\%&9.60\%&7.37\%&5.98\%&5.02\%&4.31\% \\
\hline
\end{tabular}
\end{center}
\caption{\label{tab-s-1} The integrated cross sections of $e^+e^-\to H\gamma$ in a variety range of CM energy, at both LO
and NLO accuracy. We also enumerate the corresponding values of the scalar form factors $T_{\gamma,5}$ in
(\ref{eq-c-1:decomposition}). For simplicity, we have eliminated the $\alpha$-dependence by redefining
$\widetilde{T}_{\gamma,5}\equiv \alpha^{-3/2} \,T_{\gamma,5}$.
}
\end{table*}

In Table~\ref{tab-s-1} we enumerate the predicted cross sections over a wide range of CM energy, at both LO and NLO accuracy.
In addition, we also tabulate the values of the scalar form factors $T_{\gamma,5}$ at various CM
energies. At CEPC energy range, $\sqrt{s}\approx 250$ GeV, the NLO QCD corrections have a negligible effect.
In contrast, if the \textsf{ILC} and \textsf{FCC-ee} operate at $\sqrt{s}\approx 400$ GeV,
the magnitude of the NLO QCD corrections could reach roughly $20\%$ of the LO cross section.

As can be noticed in Table~\ref{tab-s-1}, as $\sqrt{s}>2m_t$, the two-loop form factor $T_{\gamma,5}$ develops an
imaginary part due to the opening of the $t\bar{t}$ threshold.
Through an exhaustive threshold scan, we observe that the real part of $T_{\gamma,5}$ develops a logarithmic
singularity $\propto \ln|\beta|$ ($\beta\equiv \sqrt{1-{4m_t^2\over s}}$) in the vicinity of the threshold,
while the imaginary part approaches a constant above from the  $t\bar{t}$ threshold.
This is a clear sign that the nearly-on-shell $t\bar{t}$ pair in the loop carries the dominant quantum number of
${}^3S_1$.
It is well-known in such a case, the Coulomb singularity $\propto {1\over \beta}$ will arise for even higher-order diagrams
containing more Coulomb gluon exchange, so that fixed-order QCD perturbative expansion inevitably breaks down near the threshold.
For inclusive $t\bar{t}$ production near the threshold, a vast amount of literatures have been devoted to providing a
systematic description in the threshold region, by resumming Coulomb singularities to all orders and incorporating the finite
$t$ width effect, in the context of non-relativistic effective field theory NRQCD~\cite{Fadin:1987wz,Fadin:1988fn,Strassler:1990nw,Hoang:2000yr,Beneke:2015kwa}.
For our exclusive $H+\gamma$ process, one may closely follow the ansatz adopted for the pseudosalar Higgs decay into two photons when the pseudoscalar Higgs mass is about twice the top quark mass, where a $t\bar{t}$ pair is
in the ${}^1S_0$ state in the loop~\cite{Drees:1989du,Melnikov:1994jb}. The resummation of the Coulomb gluon ladder diagrams, as well as including top quark width, could also be readily fulfilled, so that one can obtain reliable predictions
for $\sigma(H\gamma)$ in the vicinity of $\sqrt{s}=2m_t$. Such a ``nonperturbative" treatment is beyond the scope of this Letter,
and we plan to present a comprehensive threshold analysis for this process in the near future.
In any rate, our fixed-order predictions recorded in Table~\ref{tab-s-1}, should still be viewed as trustworthy,
as long as $\sqrt{s}$ does not lie in proximity to the $t\bar{t}$ threshold.

%-------------------------------------------------
\begin{figure}[ht]
\begin{center}
\includegraphics*[scale=0.8]{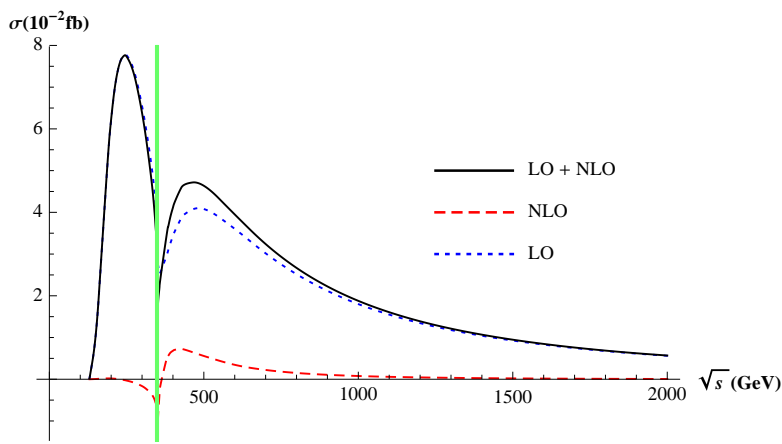}
\caption{The total cross section as a function of $\sqrt{s}$, both at LO and NLO in $\alpha_s$.
The vertical band with $\sqrt{s}=2m_t\pm 5$ GeV signifies the threshold region, inside which the
perturbative expansion is expected to break down and our fixed-order predictions become invalid.}
\label{fig-cs-s}
\end{center}
\end{figure}
%-------------------------------------------------

For the sake of clarity, in Figure~\ref{fig-cs-s} we also plot the integrated cross section as a function of the CM energy,
explicitly including the NLO QCD corrections.
In most energy ranges, the impact of the NLO QCD corrections seems rather modest,
especially when $\sqrt{s}<290$ GeV, with a relative effect typically less than $1\%$.
Nevertheless, with the increasing CM energy, the QCD corrections appear to become more pronounced,
and reach the maximum about 20\% at $\sqrt{s}\approx 400$ GeV. The effect of NLO corrections
diminishes again as the CM energy further increases, completely negligible after $\sqrt{s}> 1$ TeV.
This can be readily understood in light of the preceding discussions:
due to the dominance of the electroweak
box diagrams, the LO cross section exhibits a mild $1/s$
asymptotic scaling, while the NLO QCD corrections fall with a much faster pace
$\propto 1/s^2$, since the corresponding diagrams possess only $s$-channel topology.

%-------------------------
\setlength{\tabcolsep}{2pt}
%-------------------------
\begin{table}[hbt]
\begin{center}
\begin{tabular}{|c|c|c|c|c|c|c|c|c|c|c|c|c|c|c|}
\hline
$m_H({\rm GeV})$&40&60&80&100&120&140&160&180&200&220\\
\hline
{\rm $\sigma^{{\rm LO}}$ ($10^{-2}$ fb)}&$15.46$&$14.15$&$12.43$&$10.42$&$8.293$&$6.288$&$5.454$&$3.212$&$1.170$&$0.1659$\\
\hline
$\tilde{T}_{\gamma,5}$($10^{-4}$GeV$^{-1}$)&-3.70&0.101&5.59&12.9&22.3&34.2&48.9&67.1&89.7&118\\
%\hline
%$\tilde{T}_{Z,5} $($10^{-4}$GeV$^{-1}$)&1.35&$-0.04$&-2.04&-4.72&-8.46&-12.5&-17.9&-24.5&-32.8&-43.1\\
\hline
$\sigma^{{\rm NLO}}/\sigma^{{\rm LO}}$&0.03\%& -0.001\%& -0.05\%& -0.11\%& -0.19\%& -0.27\%& -0.33\%& -0.40\%& -0.48\%& -0.58\%\\
\hline
\end{tabular}
\end{center}
\caption{\label{tab-s-3}Cross sections of $e^+e^-\to H \gamma$ as a function of Higgs boson mass, with $\sqrt{s}$
fixed at 240 GeV.}
\end{table}
%------------------------------------

Finally in Table~\ref{tab-s-3}, we further study the dependence of the NLO QCD corrections on the Higgs boson mass,
with $\sqrt{s}$ held at 240 GeV.
The LO cross section decreases monotonously with the rising $m_H$, partly due to the shrinking phase space.
The magnitude of the NLO QCD corrections varies with $m_H$ with a modest pace,
nevertheless always less significant than $1\%$.

\section{Summary and Outlook\label{sec-summary}}

The LO predictions for the production rate of $e^+e^-\to H\gamma$
in SM are rather small at the next-generation $e^+e^-$ colliders, which
may render it a sensitive probe for the elusive beyond-SM signals.
An accurate account of this process within SM appears mandatory, if one wishes to
confidently confront the potential new physics signals in the future.
In this Letter, we have, for the first time, calculated the NLO QCD corrections to this associated $H+\gamma$
production process, and conducted a comprehensive numerical study covering a wide range of CM energy.
The key finding is that, the NLO QCD corrections to this process at CEPC can be safely neglected,
but may have sizable impact at ILC energy range. Nevertheless, the $e^+e^-\to H\gamma$ process appears
to possess the maximal production rate, $\sigma_{\rm max}\approx 0.08$ fb, around $\sqrt{s}= 250$ GeV,
therefore CEPC appears to be the ideal place to look for this rare Higgs production process.

It may also be valuable to reanalyze this process for polarized electron beam,
as is useful for the ILC experiment.

As was mentioned before, in order to obtain a precise prediction for the $e^+e^-\to H\gamma$ process,
it seems mandatory to lessen the large ambiguity inherent in the choice of the electroweak coupling constants.
This symptom can only be relieved by further including the NLO weak corrections.
As was witnessed in the Higgsstrahlung process,
incorporating the NLO weak corrections does significantly stabilize the predicted production rate for $HZ$~\cite{Denner:1992bc,Sun:2016bel}.
This is particularly important for our process, since the ambiguity due to the choice of $\alpha$ might
even bring forth greater uncertainty than the magnitude of NLO QCD corrections per se.
Despite its daunting difficulty, we hope that future fulfillment of the NLO weak corrections to this
process will significantly reduce the $\alpha$-scheme dependence.
Combined with the NLO QCD corrections calculated in this Letter,
we will then be able to provide a fairly reliable guidance for the future experimental search of
this rare Higgs production channel.

Finally, we think it is also of some interest to conduct a full-fledged study for $\sigma(H\gamma)$ in the
vicinity of the $t\bar{t}$ threshold, which necessitates to include resummation of the Coulomb gluon ladder diagrams
and top quark width effect.

%\begin{appendix}
\section*{Appendix: The compact expressions for $C^{\pm(0)}_{1,2}$\label{app-1}}

For reader's convenience, here we list the compact expressions for the LO coefficients
functions $C^{\pm(0)}_{1,2}$ appearing in (\ref{eq-cs-f}), with $m_e$ set to zero at the outset.
They can be split into three parts:
%------------------------
\begin{equation}
%------------------------
C^{\pm(0)}_{i}=\alpha^2\bigg[\sum_{V=Z,\gamma}\frac{g^{\mp}_{Ve}M_W}{2s_W(s-M_V^2)}C^{\pm(0,{\rm vert})}_{Vi}
-\frac{M_W}{s_W^3} C^{\pm(0,{\rm Wbox})}_{i}+\frac{2M_W}{s_Wc_W^2} C^{\pm(0,{\rm Zbox})}_{i}\bigg],
%------------------------
\end{equation}
%------------------------
where $i=1,2$, and
%------------------------
\begin{subequations}
%------------------------
\bqa
%------------------------
%\begin{split}
C^{\pm(0,\text{vert})}_{Vi}&&=\frac{4}{3}\frac{m_t^2}{M_W^2} N_c \left(g_{Vt}^-+g_{Vt}^+\right) \left(C_0^a+4C_2^a+4C_{12}^a+4C_{22}^a\right)\nn\\
&&+4\frac{m_H^2}{M_W^2} c_{V2}\left(C_2^b+C_{12}^b+C_{22}^b\right)+c_{V1}\left(10 C_0^b-2 C_1^b+17 C_2^b+19 C_{12}^b+19 C_{22}^b\right)\\
&&+2c_{V2}\left(2 C_0^b+2 C_1^b+3 C_2^b+C_{12}^b+C_{22}^b\right)
+c_{V3}\left(2 C_0^b-2 C_1^b+C_2^b+3 C_{12}^b+3 C_{22}^b\right),\phantom{xxxx}\nn\\
&&C^{+(0,{\rm Wbox})}_{1}=D_2^a+D_{23}^a+D_2^b-D_{23}^b-D_{33}^b,\\
&&C^{+(0,{\rm Wbox})}_{2}=D_1^a+D_1^b+D_{13}^b-D_{13}^a-D_{33}^a,\\
&&C^{-(0,{\rm Wbox})}_{1,2}=0,\\
&&C^{\pm(0,{\rm Zbox})}_{1}=-g_e^{\mp 2}\left(D_{13}^c+D_{33}^c\right),\\
&&C^{\pm(0,{\rm Zbox})}_{2}=g_e^{\mp 2}\left(D_2^c+D_{12}^c+D_{23}^c\right),
%\end{split}
%------------------------
\eqa
%------------------------
\end{subequations}
%------------------------
where $g^\pm_{Zf}=g^\pm_f, g^\pm_{\gamma f}=-Q_f$, and we define $c_{Z1}\equiv-c_W/s_W$, $c_{Z2}\equiv(s_W^2-c_W^2)/(2s_Wc_W)$, $c_{Z3}\equiv s_W/c_W$, $c_{\gamma1}=c_{\gamma2}=c_{\gamma3}\equiv1$, and
%------------------------
\begin{subequations}
\label{PV:scalar:functions}
%------------------------
\bqa
%------------------------
%\begin{split}
C^{a}_{ij\ldots}&\equiv& C_{ij\ldots}(0,s,m_H^2,m_t^2,m_t^2,m_t^2),\\
C_{ij\ldots}^{b}&\equiv& C_{ij\ldots}(0,s,m_H^2,M_W^2,M_W^2,M_W^2),\\
D_{ij\ldots}^{a}&\equiv& D_{ij\ldots}(0,s,m_H^2,u,0,0,0,M_W^2,M_W^2,M_W^2),\\
D_{ij\ldots}^{b}&\equiv& D_{ij\ldots}(0,s,0,t,0,m_H^2,0,M_W^2,M_W^2,M_W^2),\\
%------------------------
D_{ij\ldots}^{c}&\equiv& D_{ij\ldots}(0,u,m_H^2,t,0,0,0,0,M_Z^2,M_Z^2). \label{Coll:div:potential}
%\end{split}
%------------------------
\eqa
%------------------------
\end{subequations}
%----------------------
We use the same conventions for the Passarino-Veltman scalar functions as in most modern literatures (see, for instance, Ref.~\cite{Denner:2016kdg}). Note that the functions (\ref{Coll:div:potential}), which originate from
the last diagram in Figure~\ref{fig-feyn-LO}, are by themselves free from any collinear singularity.
However, if the IBP reduction is further implemented to (\ref{Coll:div:potential}),
one will end up with a linear combination of a number of
one-loop scalar MIs, some of which are IR divergent. Fortunately, one can use \textsf{LoopTools}~\cite{Hahn:1998yk} and \textsf{Collier}~\cite{Denner:2016kdg} to directly calculate these Passarino-Veltman functions in (\ref{PV:scalar:functions}).

%\end{appendix}
\vspace{0.2 cm}
%\begin{acknowledgments}
\section*{Acknowledgments}
%{\noindent\it \color{blue} Acknowledgment.}
%-----------------------
We thank Seddigheh Tizchang for the inquiry that helps us to correct some errors
in the Appendix in earlier version of the manuscript.
%-----------------------
W.-L.~S. and Q.-F.~S. wish to thank Theory Division of IHEP for the warm hospitality,
where this work was being finalized.
%-----------------------
W.-L.~S. is supported by the National Natural Science Foundation of China under Grants No.~11447031 and No.~11605144, by the Natural
Science Foundation of ChongQing under Grant No. cstc2014jcyjA00029,
and also by the Fundamental Research Funds for the Central Universities under Grant No. XDJK2016C067.
%-----------------------
The work of W.~C. and Y.~J. is supported in part by the National Natural Science Foundation of China under Grants
No.~11475188, No.~11261130311, No.~11621131001 (CRC110 by DGF and NSFC), by the IHEP Innovation Grant under contract number Y4545170Y2,
and by the State Key Lab for Electronics and Particle Detectors.
%-----------------------
The work of F.~F. is supported by the National Natural Science Foundation of China under Grant No.~11505285,
and by the Fundamental Research Funds for the Central Universities.
%-----------------------
Q.-F.~S. is supported by the National Natural Science Foundation of China under Grant No. 11375168 and No.~11475188.
%-----------------------
The Feynman diagrams in this paper were prepared using \textsf{JaxoDraw}~\cite{Binosi:2003yf,Binosi:2008ig}.
%-----------------------
We thank Seddigheh Tizchang for the inquiry that helps us to correct some mistakes in the Appendix in earlier version of the draft.
%-----------------------
%\end{acknowledgments}

\end{document}